\documentclass[letterpaper,twocolumn,10pt]{article}
\usepackage{zhanggroup}

\usepackage[ruled,vlined]{algorithm2e}
\usepackage{microtype}
\usepackage{inconsolata}
\usepackage{graphicx}
\usepackage{float}
\usepackage{makecell}
\usepackage{multirow}
\usepackage{mdframed}
\usepackage{amsmath,amssymb}
\usepackage{xspace,xparse}
\usepackage{latexsym}

\newcommand{\mypara}[1]{\noindent{\bf {#1}.} \xspace}

\newcommand{\agentname}{GuardrailAgent\xspace}
\newcommand{\inputagent}{InputAgent\xspace}
\newcommand{\outputagent}{OutputAgent\xspace}
\newcommand{\dsro}{$DSR_{overall}$\xspace}
\newcommand{\dsri}{$DSR_{input}$\xspace}
\newcommand{\foneo}{$F1_{overall}$\xspace}
\newcommand{\fonei}{$F1_{input}$\xspace}

\begin{document}
%-------------------------------------------------------------------------------

\date{}

\title{\bf GuardrailAgent: An Agentic Framework for Jailbreak Defense}

\author{
Tianze Chang\ \ \
Junjie Chu\ \ \
Yixin Wu\ \ \
Michael Backes\ \ \
Yang Zhang\thanks{Corresponding author.}
\\
\\
\textit{CISPA Helmholtz Center for Information Security}
\\
\texttt{\{tianze.chang, junjie.chu, yixin.wu, zhang\}@cispa.de, backes@acm.org}
}

\maketitle

%-------------------------------------------------------------------------------
\begin{abstract}

Despite undergoing safety training to align with human values, large language models (LLMs) remain vulnerable to jailbreak attacks that circumvent built‑in safeguards and elicit malicious behaviors given by the adversary.
Existing defenses are mostly static and fail to adapt to new or unknown attack strategies, providing limited resilience.
To address these, we propose a multi‑agent defense framework, \agentname, that offers adaptive, end‑to‑end protection.
The system autonomously gathers and summarizes attack characteristics into a continuously updated knowledge base that supports both input‑ and output‑side detection.
A feedback module integrates misclassified cases for iterative improvement at minimal cost.
Tested on over ten representative jailbreaks, the framework achieved a 93.1\% input‑side detection rate, 93.9\% overall detection, and a 95.8\% F1 score, while rapidly adapting to unseen attacks.
These results demonstrate \agentname's outstanding capability to maintain LLM safety in the face of evolving jailbreak tactics.

\end{abstract}
%-------------------------------------------------------------------------------

%-------------------------------------------------------------------------------
\section{Introduction}
%-------------------------------------------------------------------------------

Jailbreak attacks against large language models (LLMs) are rapidly evolving, with new variants emerging almost daily~\cite{MZKNASK23,CSBZ24,DLLWZLWZL24,YLYX23,CRDHPW25,CLYSBZ25,ZLMLW26}.
In contrast, current defenses, whether they involve extensive safety fine-tuning~\cite{BJNACDDFGHJKKCEEHHHJKLNOABCMOMK22,ABCDGHJJMDEHHKNOABCMOK21}) or external safeguards~\cite{AK23,JSWSKCGSGG23,MZANLAJW23,KASLFL23,IUCRIMTHFTK23}, remain largely static and non-adaptive.
They are typically optimized against previously known prompts or attack strategies, but struggle to generalize to unseen jailbreaks in a timely manner.

This mismatch severely limits practical utility.
More fundamentally, existing defenses fail to evolve in step with adversaries: their detection logic is typically tied to fixed rules, datasets, or training distributions, such as~\cite{AK23,IUCRIMTHFTK23,RWHP23}, which quickly become outdated as new jailbreak variants emerge~\cite{YLSCHSXL24,MCLYZ25}.
As a result, even defenses that are effective against known attacks often struggle to maintain robustness as jailbreak behaviors continue to evolve over time.
These limitations highlight the urgent need for a defense framework that can continuously and autonomously adapt to a rapidly shifting jailbreak landscape.

\begin{figure*}[t]
\begin{center}
\includegraphics[width=0.85\textwidth]{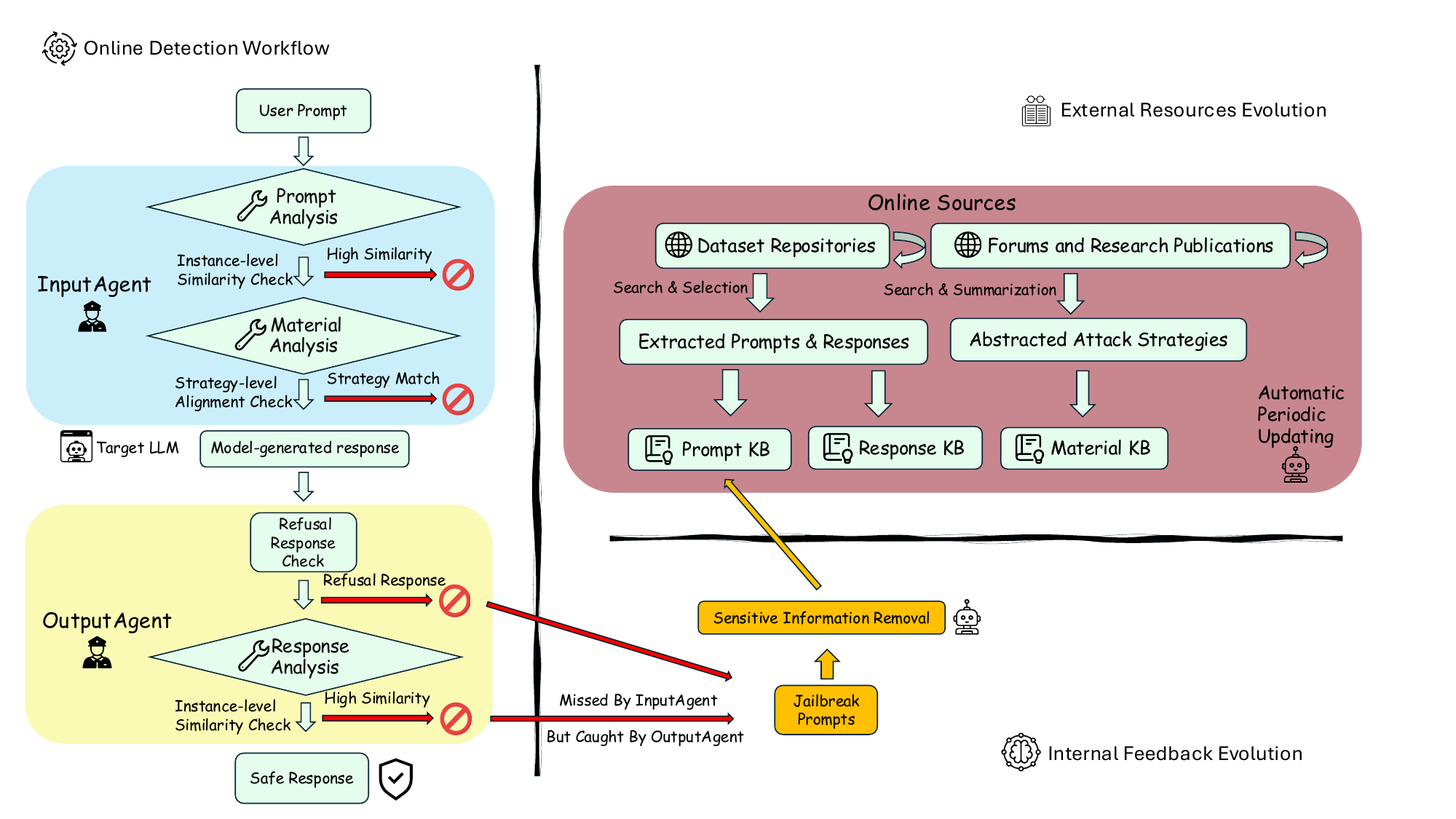}
\end{center}
\caption{Overview of \textbf{\agentname}.
The framework adopts a two-stage online detection workflow with an InputAgent and an OutputAgent for analyzing user prompts and model-generated responses.
The three knowledge bases are continuously updated through external resource evolution and an internal feedback mechanism.}
\label{fig:system}
\end{figure*}

To this end, we propose \agentname, a framework that provides comprehensive and self-evolving protection against unseen and emerging jailbreak attacks.
Unlike existing defenses that primarily rely on static patterns, \agentname is designed to continuously evolve through automated updates from both external resources and internal feedback.
This self-evolving capability enables the framework to adapt over time, allowing it to remain effective as jailbreak attacks continue to change.
As illustrated in~\autoref{fig:system}, \agentname adopts a multi-agent architecture composed of two specialized agents: \inputagent and \outputagent.
These agents conduct analysis via a knowledge-driven system, which is built upon three specialized knowledge bases (KBs) and their corresponding analysis tools: the Prompt KB and Material KB for input-side detection, and the Response KB for response monitoring.
Externally, the framework autonomously crawls the latest in-the-wild prompts and extracts and summarizes new attack strategies from relevant research papers and write-ups to regularly update its KBs.
Internally, it includes a crucial feedback loop: whenever an attack bypasses input detection but is successfully caught at the output side, the missed instance is promoted into the Prompt KB for future detection.
This mechanism ensures that the framework can autonomously learn from its past failures in a timely manner, thereby improving its overall generalization.

We conduct extensive evaluations of \agentname on three popular LLMs as target models, using a benchmark suite of jailbreak prompts containing more than ten representative jailbreak attack strategies.
To assess the balance between detection performance and model utility, we also evaluate the models on benign queries to ensure that their utility is not compromised.
Overall, it consistently surpasses all baseline defenses, achieving the highest detection rates in both input-side and overall evaluations while preserving model utility.
For example, when targeting the \texttt{ChatGLM3-6B} model, it achieves 0.939 \dsro and 0.958 \foneo, outperforming baselines, e.g., OpenAI Moderation~\cite{OpenAI_Moderation2} and Llama Guard 4~\cite{IUCRIMTHFTK23,Llama_Guard_2,llama_guard_3,llama_guard4}, exceeding them by at least 0.219 \dsro and 0.123 \foneo.
We further evaluate \outputagent on jailbreak prompts that manage to bypass the input detector and trigger unsafe responses from the LLM, and show that it consistently outperforms baseline approaches.
Additionally, we study the contribution of its dynamic updating mechanisms, revealing that both continuous updates from external resources and updates informed by internal feedback enhance detection performance.
For example, in feedback mode, we observe that the uninitialized \agentname, after processing on average 90 adversarial prompts, successfully adapts and improves its detection performance from 0.033 to 0.800.

Our contributions are summarized as follows:
\begin{itemize}
    \item We propose \textbf{\agentname}, a multi-agent defense framework that integrates input- and output-side detection with dynamically evolving knowledge bases.
    It autonomously collects jailbreak prompts, literature, and community write-ups, and continuously updates itself through web crawling and feedback-driven refinement.
    \item Our extensive evaluations on three popular LLMs with over 1,400 harmful queries and ten representative jailbreak strategies show that \agentname achieves a 0.931 input-side detection rate, a 0.939 overall detection rate, and a 0.958 F1 score on benign queries, substantially outperforming existing baselines such as OpenAI Moderation and Llama Guard 4.
    \item We further demonstrate \agentname’s strong adaptivity: in feedback mode, the system quickly strengthens itself by promoting missed instances into its knowledge bases, improving its average detection performance to 0.800 after only 90 adversarial prompts.
\end{itemize}

%-------------------------------------------------------------------------------
\section{Related Work}
%-------------------------------------------------------------------------------

%-------------------------------------------------------------------------------
\subsection{Jailbreak Attacks}
\label{section:jailbreak_attack}
%-------------------------------------------------------------------------------

Jailbreak attacks aim to bypass built-in safety mechanisms of LLMs by crafting adversarial prompts that induce restricted or harmful outputs.
Such prompts are widely shared in community forums and red-teaming efforts, and have been aggregated into public datasets such as \emph{do-anything-now}~\cite{SCBSZ24} and \emph{WildJailbreak}~\cite{JRHEBKMLSCD24}.
Many jailbreaks exploit predictable model behaviors through fixed prompt structures or role-play instructions.
Representative examples include \emph{AIM} and \emph{Devmode}~\cite{SCBSZ24}, where attackers instruct models to adopt unrestricted personas or override safety constraints.
Others rely on obfuscation techniques, such as unconventional encodings (e.g., \texttt{base64}~\cite{RVNAC23}), to conceal malicious intent from simple filters.
Beyond handcrafted prompts, adversaries increasingly employ automated search and mutation strategies to discover effective jailbreak variants.
For instance, \emph{GCG}-style approaches~\cite{ZWCNKF23,LLLLCH24} iteratively refine prompt suffixes to bypass safeguards, while \emph{PAIR}-style methods~\cite{CRDHPW25,MZKNASK23} evolve prompt--response pairs that elicit unsafe outputs.

%-------------------------------------------------------------------------------
\subsection{Jailbreak Defenses}
\label{section:jailbreak_defense}
%-------------------------------------------------------------------------------

A variety of defenses have been proposed to mitigate jailbreak attacks, aiming either to prevent unsafe behavior or to detect adversarial attempts before harmful content is generated.
Some approaches strengthen model alignment through specially crafted instructions or system prompts, encouraging the model to remain safe even under adversarial inputs, as demonstrated by methods like \emph{Self-Reminder}~\cite{XYSCLCXW23}.
Others focus on transforming or sanitizing potentially risky queries, for example, through \emph{paraphrasing}~\cite{JSWSKCGSGG23} or \emph{retokenization}~\cite{JSWSKCGSGG23}, which seek to neutralize malicious intent while preserving the original semantics.
In parallel, detection-oriented strategies employ classifiers or filters to identify unsafe prompts or outputs, such as the \emph{OpenAI Moderation}~\cite{OpenAI_Moderation2} or the fine-tuned \emph{Llama Guard}~\cite{IUCRIMTHFTK23,Llama_Guard_2,llama_guard_3,llama_guard4}.

While these defenses offer useful coverage for known threats, they typically rely on static rules or limited training data, and thus struggle to generalize to newly emerging jailbreak techniques.
Specifically, instruction- or prompt-based alignment methods rely on fixed safety instructions that remain static after deployment, making them vulnerable to newly crafted or adaptive jailbreak prompts;
and detection-oriented strategies based on classifiers or filters are trained on limited datasets and often exhibit out-of-distribution generalization issues when facing novel or evolving jailbreak techniques.

%-------------------------------------------------------------------------------
\section{\agentname}
\label{section:agent}
%-------------------------------------------------------------------------------

%-------------------------------------------------------------------------------
\subsection{Problem Statement}
%-------------------------------------------------------------------------------

We assume a powerful and unconstrained adversary capable of operating with full black-box or white-box access to the target LLM.
In contrast, our proposed defense operates strictly under a black-box access assumption.
Thus, it relies exclusively on analyzing the input prompt and the resulting output response, thus ensuring the solution is broadly applicable to closed-source, proprietary LLMs.

\mypara{Design Goals}
\begin{itemize}

    \item \textbf{Dynamic and Autonomous Adaptation.}
    The defense should self-evolve to remain robust against unseen or novel jailbreak strategies.

    \item \textbf{Holistic and Effective Defense.}
    The defense should achieve strong detection performance across diverse attack families while preserving model utility, ensuring that benign interactions remain largely unaffected.

\end{itemize}

%-------------------------------------------------------------------------------
\subsection{Framework Design}
\label{section:framework}
%-------------------------------------------------------------------------------

\subsubsection{Detection Workflow}

As illustrated in~\autoref{fig:system}, the framework operates a two-stage defense.
On the input side, an input agent leverages Prompt and Material Knowledge Bases (KBs), which are continuously updated from external resources like web-sourced jailbreak prompts and technical documents, to analyze and block malicious user requests.
Prompts that pass this check elicit an LLM response, which is then scrutinized by an output agent with a Response Knowledge Base to detect harmful content.
Additionally, an internal feedback loop captures any attacks missed by \inputagent and feeds them back into the knowledge bases, enabling the system to self-improve and maintain robust defenses against constantly evolving attack strategies.

\subsubsection{Components}

\mypara{Knowledge Bases}
Below is the introduction to the three different knowledge bases.
\begin{itemize}

    \item \textbf{Prompt KB.}
    The Prompt Knowledge Base stores concrete jailbreak prompt instances collected from public datasets and in-the-wild sources.
    It enables fast instance-level detection via pattern matching and semantic similarity, allowing \inputagent to precisely identify previously observed attack vectors.

    \item \textbf{Material KB.}
    The Material Knowledge Base contains structured summaries of jailbreak strategies distilled from academic papers and technical reports.
    Unlike the Prompt KB, it captures attack patterns at the strategy level, enabling \inputagent to generalize beyond exact prompt formulations and detect novel jailbreaks that follow known techniques.

    \item \textbf{Response KB.}
    The Response Knowledge Base maintains representative non-compliant model outputs associated with successful jailbreaks.
    It encodes characteristic linguistic and structural patterns of compromised responses and is used by \outputagent to identify harmful outputs that bypass input-side defenses.

\end{itemize}
These KBs are built and updated from both external resources and internal feedback (see~\autoref{section:evolve_comp}).

\mypara{\inputagent}
The agent evaluates whether a given prompt $x$ resembles known jailbreak patterns using two tools:
(i) the \texttt{Prompt Analysis Tool}, which compares $x$ against entries in the Prompt KB and returns a similarity score based on the top-3 most similar jailbreak prompts; and
(ii) the \texttt{Material Analysis Tool}, which leverages the Material KB to assess alignment with documented jailbreak strategies.
The \inputagent adopts a cascaded decision process: prompts that closely match known jailbreak instances are first filtered and flagged using the Prompt Analysis Tool, while remaining prompts are further examined by the Material Analysis Tool.
If $x$ is flagged at any stage, the system outputs a refusal; otherwise, the prompt is forwarded to the target model $\mathcal{M}$.

\mypara{\outputagent}
The output detection agent evaluates whether a model-generated response resembles any known jailbreak response patterns.
The underlying intuition is that, similar to input prompts, responses induced by jailbreak attempts often exhibit similar patterns, such as accepting malicious roles or providing step-by-step instructions for forbidden scenarios.
As a first step, it checks whether the target model has already produced a refusal; if so, this refusal is directly returned as the final decision.
Otherwise, the agent invokes the \texttt{Response Analysis Tool}, which compares the response against entries in the Response KB and returns a similarity score indicating the likelihood of a jailbreak.
This score is then used to determine whether the response should be blocked.

The end-to-end detection process is summarized in the pseudocode in~\autoref{appendix:pseudocode}.

\subsubsection{Dynamic Evolving Mechanisms}
\label{section:evolve_comp}

A central design goal of \agentname is to remain effective against newly emerging and previously unseen jailbreak strategies.
To this end, we design a fully automated knowledge construction and maintenance pipeline that continuously updates all three knowledge bases through external evolving and internal feedback, as illustrated in ~\autoref{fig:system}.
Unlike static defenses that rely on fixed datasets or manually curated rules, \agentname autonomously expands and maintains its knowledge bases over time without human intervention.

\paragraph{Evolving from External Resources.}
The external evolving component is responsible for periodically discovering newly available jailbreak-related resources and incorporating them into the knowledge bases.
This process runs at a user-configurable frequency (e.g., weekly or monthly), allowing the system to adapt its update cadence to the pace at which new jailbreak behaviors emerge.

For the Prompt KB and Response KB, the system searches for jailbreak-related datasets based on their metadata and selects representative resources according to lightweight popularity signals, which are used to approximate dataset quality and community adoption.
From the selected datasets, jailbreak prompts and responses are extracted through simple field matching, requiring no dataset-specific customization.
In our experiments, the Prompt KB and Response KB are constructed from widely used jailbreak datasets, which provide reasonably broad and representative coverage of jailbreak behaviors encountered in practice.

For the Material KB, the system periodically searches for newly published research works related to jailbreak attacks based on their titles and abstracts.
Each retrieved work is summarized into an attack-strategy entry using an automated summarization process designed to extract jailbreak methods and characteristic patterns.
The Material KB is maintained exclusively through this external evolving process, enabling the system to continuously incorporate newly proposed jailbreak strategies.

\mypara{Evolving from Internal Feedback}
The agent also evolves dynamically based on its internal feedback.
Specifically, it incorporates a feedback loop: whenever the input detection agent fails to identify a jailbreak but the output detection agent successfully catches it, the misclassified prompt is stored as a candidate entry for future updates to the Prompt KB.
This mechanism allows the detection agent to quickly achieve robust performance against previously unseen attack strategies.

%-------------------------------------------------------------------------------
\section{Evaluation}
\label{section:evaluation}
%-------------------------------------------------------------------------------

%-------------------------------------------------------------------------------
\subsection{Experiment Settings}
\label{section:Experiment Settings}
%-------------------------------------------------------------------------------

\mypara{Target Models}
We select three models with relatively weak internal safety protection as the target of jailbreak attacks, including \texttt{ChatGLM3-6B}~\cite{ChatGLM3}, \texttt{Vicuna-7B}~\cite{Vicuna}, and \texttt{GPT-3.5-turbo}~\cite{chatgpt}, which allows us to more clearly assess the effectiveness of the external guardrail.

\mypara{GuardrailAgent Settings}
\agentname is built upon the SmolAgents framework.
In our main experiments, all agents are driven by GPT-4.1-mini~\cite{O23}.
For the ablation study, we also consider Gemini-2.5-flash~\cite{T23} as the base model to drive agents.
All target models and base models are evaluated using their default settings.
All reported results are obtained with a warmed-up \agentname, where the knowledge bases are initialized through an initial dynamic evolution phase (details in~\autoref{appendix:settings}).

\mypara{Baseline Defenses}
We consider four representative baseline defenses.

 \texttt{Llama Guard 4}~\cite{llama_guard4}, an alignment-tuned model providing dual-sided protection for prompts and responses;
 \texttt{OpenAI Moderation}~\cite{OpenAI_Moderation2}, a widely deployed content moderation system operating on both inputs and outputs;
 and \texttt{PromptGuard}~\cite{prompt-guard}, a lightweight model specifically designed for input-side filtering.

In addition, we include \texttt{AutoDefense}~\cite{ZWZWW24}, a multi-agent defense framework that uses role-specialized LLM agents to collaboratively filter harmful responses, serving as a representative agent-based baseline.
Since AutoDefense operates exclusively on the output side and does not incorporate any learning mechanisms, we compare it with \agentname when evaluating the OutputAgent in isolation.
All baselines are evaluated under their default settings.

\mypara{Evaluation Metrics}
We evaluate \agentname at both the input stage and the end-to-end system level.
We report detection success rates \dsri and \dsro, where \dsri measures the fraction of jailbreak prompts detected at the input stage, and \dsro counts an instance as detected if either its prompt or response is flagged.
In addition, we report \fonei and \foneo to capture the trade-off between detection performance and benign utility.
To address class imbalance, we use weighted F1 scores, where class-wise F1 scores are averaged according to their sample proportions (formal definition in~\autoref{appendix:definition}).

\begin{table*}[!t]
\centering
\scalebox{0.65}{
\setlength{\tabcolsep}{3pt}
\begin{tabular}{c|cc|cc|cc|cc|cc|cc}
\toprule
\multirow{2}{*}{\textbf{Defense}}
  & \multicolumn{2}{c|}{\textbf{DSR (ChatGLM3-6B)}}
  & \multicolumn{2}{c|}{\textbf{DSR (Vicuna-7B)}}
  & \multicolumn{2}{c|}{\textbf{DSR (GPT-3.5-turbo)}}
  & \multicolumn{2}{c|}{\textbf{F1 (ChatGLM3-6B)}}
  & \multicolumn{2}{c|}{\textbf{F1 (Vicuna-7B)}}
  & \multicolumn{2}{c}{\textbf{F1 (GPT-3.5-turbo)}} \\
 & \textit{Input} & \textit{Overall}
 & \textit{Input} & \textit{Overall}
 & \textit{Input} & \textit{Overall}
 & \textit{Input} & \textit{Overall}
 & \textit{Input} & \textit{Overall}
 & \textit{Input} & \textit{Overall}
 \\
\midrule
PromptGuard          & 0.535 & -- & 0.594 & -- & 0.529 & -- & 0.716 & -- & 0.755 & -- & 0.736 & -- \\
OpenAI Moderation    & 0.395 & 0.396 & 0.383 & 0.383 & 0.332 & 0.332 & 0.612 & 0.613 & 0.603 & 0.603 & 0.598 & 0.598 \\
Llama Guard 4         & 0.676 & 0.720 & 0.683 & 0.731 & 0.694 & 0.691 & 0.808 & 0.835 & 0.812 & 0.841 & 0.835 & 0.833 \\
\agentname           & \textbf{0.931} & \textbf{0.939} & \textbf{0.909} & \textbf{0.931} & \textbf{0.920} & \textbf{0.940} & \textbf{0.953} & \textbf{0.958} & \textbf{0.940} & \textbf{0.953} & \textbf{0.950} & \textbf{0.961} \\
\bottomrule
\end{tabular}
}
\caption{
Detection success rates (\dsri, \dsro) and weighted F1 scores (\fonei, \foneo) of \agentname and baselines on three target LLMs.
}
\label{tab:dsr_f1}
\end{table*}

\begin{table*}[!t]
\centering
\scalebox{0.75}{
\setlength{\tabcolsep}{3pt}
\begin{tabular}{c|ccccc|cccccc}
\toprule
\multirow{2}{*}{\textbf{Defense Method}}
 & \multicolumn{5}{c|}{\textbf{Model-Independent Attacks}}
 & \multicolumn{5}{c}{\textbf{Model-Dependent Attacks}} \\
\cmidrule(lr){2-6} \cmidrule(lr){7-11}
 & WildJailbreak
 & AIM & Base64 & Devmodev2 & Combination &  AutoDAN & GCG
 & GPTFuzz & TAP & PAIR \\
\midrule
PromptGuard & 0.438
  & \textbf{1.000} & 0.000 & \textbf{1.000} & 0.000 & \makecell{\textbf{1.000} \\ \textbf{1.000} \\ --}
  & \makecell{0.181 \\ 0.600 \\ --} & \makecell{\textbf{1.000} \\ \textbf{1.000} \\ \textbf{1.000}} & \makecell{0.287 \\ 0.331 \\ 0.375} & \makecell{0.344 \\ 0.331 \\ 0.325} \\
\midrule
OpenAI Moderation & 0.492
  & 0.552 & 0.000 & 0.404 & 0.000 & \makecell{0.512 \\ 0.506 \\ --}
  & \makecell{0.512 \\ 0.463 \\ --} & \makecell{0.644 \\ 0.644 \\ 0.606} & \makecell{0.394 \\ 0.388 \\ 0.362} & \makecell{0.544 \\ 0.494 \\ 0.400} \\
\midrule
Llama Guard 4 & 0.540
  & 0.490 & \textbf{1.000} & 0.648 & 0.967 & \makecell{0.719 \\ 0.725 \\ --}
  & \makecell{0.619 \\ 0.762 \\ --} & \makecell{0.863 \\ 0.806 \\ 0.750} & \makecell{0.506 \\ 0.556 \\ 0.494} & \makecell{0.669 \\ 0.619 \\ 0.487} \\
\midrule
\agentname & \textbf{0.875}
  & \textbf{1.000} & 0.981 & \textbf{1.000} & \textbf{0.994} & \makecell{\textbf{1.000} \\ \textbf{1.000} \\ --}
  & \makecell{\textbf{0.938} \\ \textbf{0.806} \\ --} & \makecell{\textbf{1.000} \\ \textbf{1.000} \\ \textbf{1.000} } & \makecell{\textbf{0.781} \\ \textbf{0.825} \\ \textbf{0.906}} & \makecell{\textbf{0.756} \\ \textbf{0.769} \\ \textbf{0.700}} \\
\bottomrule
\end{tabular}
}
\caption{
Overall detection success rate (\dsro) of \agentname and baseline defenses across individual jailbreak attacks.
\dsro counts an instance as detected if either its prompt or its response is identified as malicious, representing the end-to-end detection capability.
For model-dependent attacks, each cell reports results on attacks against three target LLMs: ChatGLM3-6B (top), Vicuna-7B (middle), and GPT-3.5-turbo (bottom).
}
\label{tab:dsr}
\end{table*}

\mypara{Evaluation Datasets}
It consists of both jailbreak and benign queries to measure detection performance while preserving model utility.

\noindent \textit{Jailbreak Queries.}
We evaluate \agentname using a diverse set of jailbreak attacks drawn from the JailbreakRadar~\cite{CLYSBZ25} benchmark and the WildJailbreak~\cite{JRHEBKMLSCD24} dataset, covering a set of representative jailbreak strategies. These attacks include both model-independent attacks based on static templates and obfuscation, and model-dependent attacks that require iterative optimization or target-model feedback.
Detailed descriptions of individual attack methods are provided in~\autoref{appendix:dataset}.

\noindent \textit{Benign Queries.}
To further evaluate the models' utility, we construct a benign dataset of 1,000 non-harmful prompts.
We first generate 900 safe, legal instructions using ChatGPT-5~\cite{gpt5}, then augment 100 of them into corresponding \emph{role-play variants} to reflect realistic user behavior, where role-play is common but not malicious.
The detailed generation prompt and representative examples are provided in~\autoref{appendix:benign}.

\mypara{Runtime and Token Cost Evaluation}
We characterize the average runtime latency and token consumption of \agentname to assess its computational overhead.
All measurements are conducted on a single-machine setup with a single NVIDIA GeForce RTX 4090 GPU.

%-------------------------------------------------------------------------------
\subsection{Evaluation Results}
%-------------------------------------------------------------------------------

\mypara{Main Evaluation}
We report the performance of \agentname with full components in~\autoref{tab:dsr_f1}.
It consistently surpasses all baseline defenses, achieving the highest detection rates in both input-side and overall evaluations.
For example, on ChatGLM3-6B, it achieves a detection success rate of 0.931 \dsri and 0.939 \dsro, compared with 0.720 \dsro for the strongest baseline.
Meanwhile, \agentname achieves 0.958 \foneo, while all baselines remain below 0.835 \foneo, highlighting that our agent maintains a strong balance between high detection capability and preservation of model utility.

\mypara{Attack-Wise Detection Analysis}
We conduct a detailed analysis of detection performance for each individual attack.
As shown in~\autoref{tab:dsr}, \agentname consistently outperforms baseline defenses across a wide range of jailbreak attacks.
Specifically, on the in-the-wild dataset, it surpasses the baselines by a large margin, achieving 0.875 \dsro, notably higher than 0.540 from Llama Guard 4 and 0.492 from OpenAI Moderation.
For other model-independent attacks, it also achieves remarkably high detection on attacks with static templates such as \texttt{AIM} and \texttt{DevMode}, both reaching 1.000, demonstrating the advantage of semantic similarity retrieval.
Performance remains strong across model-dependent attacks: \texttt{AutoDAN} and \texttt{GPTFuzz} both achieve 1.000 \dsro. Additionally, \texttt{GCG}, \texttt{TAP}, and \texttt{PAIR} also perform significantly better than the baselines.

\begin{table*}[!t]
\centering
\scalebox{0.75}{
\setlength{\tabcolsep}{6pt}
\begin{tabular}{c|lccc}
\toprule
\textbf{Defense Type} & \textbf{Defense Method}
& \textbf{Input Miss Rate}
& \textbf{Output Detection Rate}
& \textbf{DSR$_{\text{overall}}$} \\
\midrule
\multirow{3}{*}{Input–Output}
& OpenAI Moderation & 56.2\% & 3.1\% & 45.5\% \\
& Llama Guard 4 & 40.8\% & 13.7\% & 64.8\% \\
& \agentname & \textbf{13.7\%} & \textbf{15.9\%} & \textbf{88.5\%} \\
\midrule
Output-Only
& AutoDefense & -- & \textbf{47.3\%} & 47.3\% \\
\bottomrule
\end{tabular}
}
\caption{Output-side detection analysis on a strengthened subset consisting only of jailbreak prompts that successfully induce unsafe responses from the target model, identified with \textit{strong-reject evaluator}.
Input Miss Rate measures the fraction of prompts that bypass the input detector. Output Detection Rate measures the fraction of these missed prompts detected at the output stage. DSR$_{\text{overall}}$ captures the resulting end-to-end detection success on this subset.
}
\label{tab:output_detection}
\end{table*}

\mypara{False Positive Analysis}
On the revised benign dataset, \agentname achieves a 1.6\% false positive rate while maintaining strong jailbreak detection capability, compared with 0.6\% for PromptGuard, 2.2\% for OpenAI Moderation, and 0.1\% for Llama Guard 4.
However, further analysis shows that most of these false positive cases do not elicit unsafe or policy-violating behaviors on modern target LLMs, suggesting that the remaining false positives primarily reflect conservative retrieval-based matching.
Further details and representative examples are provided in~\autoref{appendix:fp}.

\mypara{\outputagent}
Using the full evaluation dataset does not adequately capture the effectiveness of the output agent, as many prompts never reach the output stage.
We therefore conduct a dedicated evaluation focusing on jailbreak prompts whose responses bypass the input detector and elicit unsafe content from the LLM.
To identify these prompts, we follow previous work~\cite{ZATGE24} and employ a \textit{strong-reject evaluator}~\cite{SLBTHPASEWT24} to analyze all jailbreak queries targeting on ChatGLM3-6B (see details in~\autoref{appendix:strong-reject}).

For this subset of confirmed jailbreak prompts, we measure:
(i) the proportion of jailbreak prompts that bypass the input detector, denoted as the \emph{Input Miss Rate};
(ii) the fraction of these missed prompts subsequently detected at the output stage, denoted as the \emph{Output Detection Rate}; and
(iii) \dsro as we defined in \autoref{section:Experiment Settings}.

As shown in~\autoref{tab:output_detection}, \agentname not only maintains the highest input-side detection rate, i.e., lowest miss rate, but also achieves the strongest output-side detection among the missed prompts among dual-sided defenses.
These results demonstrate that \outputagent provides effective secondary protection, capturing jailbreak responses that evade the input filter and ensuring end-to-end robustness.
AutoDefense achieves the highest output detection rate, which is expected since it does not perform any input-side filtering and instead evaluates all generated responses at the output stage. However, its end-to-end effectiveness remains limited on this strengthened subset.

\mypara{Generalizability Across Base LLMs}
To further verify \agentname's generalizability, we substitute its base LLM with \texttt{Gemini-2.5-flash}.
As shown in ~\autoref{tab:drivingmodel}, it achieves 0.856 \dsri and 0.878 \dsro, demonstrating strong detection capability without compromising model utility.
The consistent performance across distinct base architectures indicates that \agentname's framework design is inherently robust and not tied to a particular base LLM.

\mypara{Runtime and Token Cost}
GuardrailAgent involves multiple LLM calls per query. On average, the InputAgent consumes 8.5K input tokens and 1.6K output tokens with 6.8\,s latency, while the OutputAgent consumes 5.3K input tokens and 0.36K output tokens with 3.6\,s latency.
Using GPT-4.1-mini pricing, this corresponds to approximately \$0.006 when the query is classified at the InputAgent stage and \$0.009 when the pipeline proceeds to OutputAgent.
The measured latency is dominated by external API round-trip time rather than computation inside GuardrailAgent.

\begin{table}[!t]
  \centering
  \small
  \setlength{\tabcolsep}{1pt}
  \begin{tabular}{lcc}
    \toprule
    \textbf{Defense Method} & \textbf{Input-Side Time(s)} & \textbf{Output-Side Time(s)} \\
    \midrule
    GuardrailAgent & 6.8 & 3.6 \\
    AutoDefense(ex-3) & --  & 8.0 \\
    \bottomrule
  \end{tabular}
  \caption{Average runtime per query (seconds). AutoDefense is an output-only defense and therefore does not perform input-side filtering.}
  \label{tab:runtime_latency}
\end{table}

\begin{table}[!t]
  \centering
  \scalebox{0.75}{
  \setlength{\tabcolsep}{3pt}
  \begin{tabular}{l c c c c}
    \toprule
    \textbf{Base LLM} & \dsri &  \dsro & \fonei & \foneo \\
    \midrule
    GPT-4.1-mini & 0.931 & 0.939 & 0.953 & 0.958\\
    Gemini-2.5-flash & 0.856 & 0.878 & 0.909 & 0.922\\
    \bottomrule
  \end{tabular}
  }
  \caption{The performance of \agentname with different base LLM.
  The target LLM is set to ChatGLM3-6B.}
  \label{tab:drivingmodel}
\end{table}

%-------------------------------------------------------------------------------
\subsection{Dynamic Evolving Evaluation}
%-------------------------------------------------------------------------------

\begin{figure}[!t]
\begin{center}
\includegraphics[width=\columnwidth]{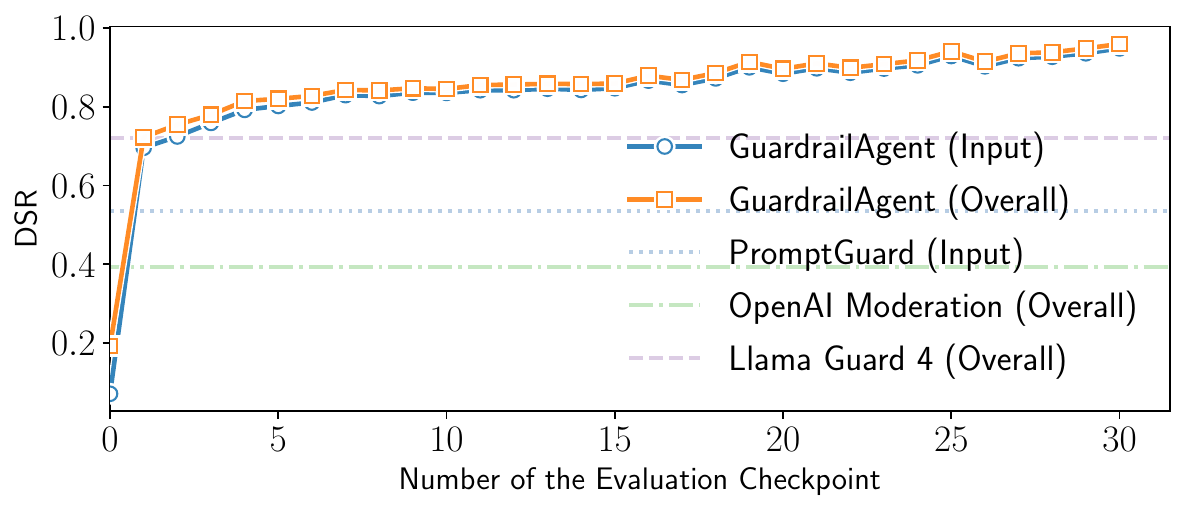}
\end{center}
\caption{DSR of \agentname and baseline defenses across 30 evaluation checkpoints of knowledge acquisition.
}
\label{fig:timeline}
\end{figure}

\begin{figure}[!t]
    \centering
    \includegraphics[width=\columnwidth]{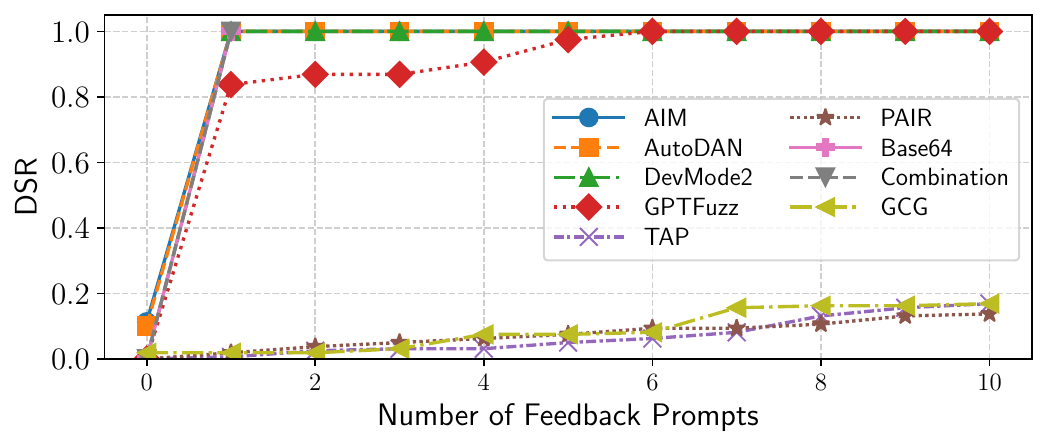}
    \caption{
    Feedback-based detection performance across jailbreak attacks against ChatGLM3-6B.
    Each subplot shows the DSR as a function of the number of feedback prompts added to the Prompt KB.
    }
    \label{fig:feedback_results}
\end{figure}

\mypara{External Evolving Component}
We next explore the effectiveness of the evolving component as it gradually incorporates external resources.
Each \textit{evaluation checkpoint} represents an intermediate state after integrating a portion of collected prompts and materials, with the process divided into 30 equally spaced checkpoints.
As shown in~\autoref{fig:timeline}, after only 1–2 checkpoints, the system reaches baseline-level performance and continues to improve steadily through subsequent checkpoints, demonstrating that knowledge accumulation consistently enhances detection capability.

\mypara{Internal Evolving Component}
To isolate the effectiveness of the internal evolving component, we disable the use of any external resources during the updating process.
The \inputagent relies solely on feedback signals provided by the output agent to iteratively refine its detection behavior.
As illustrated in~\autoref{fig:feedback_results}, we report the detection performance against nine jailbreak attacks.

For most attacks, the internal feedback mechanism proves highly effective.
For example, \texttt{AIM}, \texttt{Devmode}, \texttt{AutoDAN} and \texttt{GPTFuzz} achieve nearly 1.0 DSR after collecting only a small number of feedback examples.
This effectiveness largely stems from the fact that attacks using static templates generate final jailbreak prompts with minimal semantic variation.
For instance, different \texttt{AutoDAN} prompts may appear distinct, but they consistently convey the instruction ``role-play as ChatGLM in Developer Mode,'' resulting in strong semantic similarity captured by our retrieval process.
In contrast, detection performance is much weaker on \texttt{TAP} and \texttt{PAIR}, as both attacks iteratively refine an initial prompt through the attacker LLM's chain-of-thought reasoning combined with a static template.
This process introduces much greater semantic diversity in the resulting prompts, making them more difficult to detect.
We further conduct an experiment to identify which tools within \inputagent are most useful for detection, provided in~\autoref{appendix:tool_analysis}, and a brief discussion of how these observations relate to the internal evolving mechanism.

To further consolidate these findings, we aggregate all category-specific experiments and conduct three runs on a combined jailbreak dataset of over 1400 prompts.
Starting from an uninitialized state, the \agentname framework progressively adapts through feedback and reaches an average input-side detection rate of 0.800 after processing approximately 90 adversarial prompts.
This demonstrates that the internal evolving component alone enables substantial self-improvement even without any external knowledge sources.

%-------------------------------------------------------------------------------
\subsection{Adaptive Adversary Analysis}
\label{sec:adaptive}
%-------------------------------------------------------------------------------

We evaluate the robustness of \inputagent against two types of adaptive adversaries: prompt-injection suffix attacks and semantic paraphrasing attacks.

For prompt injection, we apply canonical prompt-injection patterns~\cite{LJGJG24} as adversarial suffixes to jailbreak prompts from the evaluation dataset, with the attacker aiming to manipulate \inputagent into producing \texttt{uncertain} outputs.
Across all evaluated injection types, \inputagent maintains strong robustness and we do not observe successful manipulations that reliably compromise the agent’s decision behavior.
Detailed settings and quantitative results are provided in~\autoref{appendix:adaptive_injection}.

We additionally evaluate adaptive semantic paraphrasing attacks targeting the Prompt KB retrieval mechanism.
Starting from jailbreak prompts targeting ChatGLM3-6B, we iteratively apply synonym substitutions using GPT-4o-mini to reduce retrieval similarity while preserving jailbreak intent.
Although these mutations increase Prompt KB retrieval bypass, the overall input-side detection rate of \agentname remains strong and even improves under adaptive paraphrasing attacks, suggesting that optimizing prompts specifically to evade retrieval similarity may simultaneously weaken semantic or structural patterns relied upon by the Material Analysis Tool.
Detailed settings and results are provided in~\autoref{appendix:semantic_adaptive}.

%-------------------------------------------------------------------------------
\section{Conclusion}
%-------------------------------------------------------------------------------

In this work, we introduced \agentname, an adaptive defense framework that provides end-to-end protection against jailbreak attacks through an agentic design.
Unlike prior defenses that rely on static rules or fixed detection logic, \agentname continuously evolves through autonomous knowledge-base construction and feedback-driven refinement, enabling robust detection on both the input and output sides.
Our experiments across more than ten representative attack families and three target models show that the system achieves a 93.1\% input-side detection rate, a 93.9\% overall detection rate, and a 95.8\% F1 score, while maintaining high usability under benign scenarios.
These results demonstrate that \agentname is both effective and practical.
We believe \agentname represents a promising step toward building resilient, self-improving safety mechanisms for large language models facing continuously evolving jailbreak threats.
Our code is publicly available at \url{https://github.com/TrustAIRLab/GuardrailAgent}.

%-------------------------------------------------------------------------------
\section*{Limitations}
%-------------------------------------------------------------------------------

While our results demonstrate the effectiveness and adaptivity of \agentname, the framework still has inherent limitations.
This work focuses on the design and evaluation of a modular multi-agent defense framework, and instantiates it with relatively lightweight components, such as similarity-based retrieval and literature-guided analysis.
While these mechanisms already cover a wide range of jailbreak scenarios, they may be less effective against attacks that exploit deeper semantic ambiguities or require more advanced reasoning capabilities.
The modular design of \agentname allows stronger analysis modules or future defensive tools to be incorporated as they become available.

%-------------------------------------------------------------------------------
\subsection*{Ethical Considerations}
%-------------------------------------------------------------------------------

This work focuses on defending large language models against jailbreak attacks, which can bypass built-in safeguards and elicit harmful or policy-violating outputs.
We believe that developing more adaptive and practical defenses contributes positively to the safe deployment of LLMs in real-world applications, reducing the risk of misuse for generating disinformation, offensive content, or malicious code.

At the same time, deploying automated guardrail systems also raises ethical considerations related to data handling, system reliability, and the broader attack–defense ecosystem.

A primary concern is data governance and user privacy, as guardrail mechanisms may involve analyzing user prompts and model responses. GuardrailAgent is therefore designed to minimize privacy risks by supporting a logging-free mode, in which no raw user prompts or responses are persistently stored. When feedback-based knowledge base updates are enabled to improve adaptivity, the system does not retain raw prompts; instead, it incorporates only anonymized, non-identifying representations, such as high-level summaries or embedding vectors. This design allows adaptivity to be improved without requiring long-term storage of sensitive user content.

Another consideration is that defensive systems are inherently imperfect. False positives may degrade user experience by incorrectly flagging benign interactions, while false negatives could still allow harmful outputs to pass through. These trade-offs highlight the importance of evaluating both safety and usability, which we explicitly address through experiments on benign prompts, including role-playing scenarios.

In addition, the defense framework itself may become a target of adversarial manipulation, for example through prompt injection attacks against coordinating agents or analysis components. Finally, as with many security mechanisms, publicizing defense strategies may indirectly inform adversaries, contributing to an ongoing arms race between attack and defense.

To mitigate these risks, we design GuardrailAgent to be modular and extensible, allowing stronger defensive tools and safeguards to be integrated as they become available. Our intent in releasing this research is to foster a deeper community understanding of jailbreak defense and to support the development of safer and more responsible LLM deployments.

Because jailbreak benchmarks inherently contain harmful, policy-violating, or offensive prompts and responses, we restrict their use to controlled research and evaluation purposes.
And all external datasets and resources used in this work are publicly available and are used solely for research purposes in accordance with their intended use and distribution terms whenever applicable.

%-------------------------------------------------------------------------------
\bibliographystyle{plain}
\bibliography{references}
%-------------------------------------------------------------------------------

\appendix

%-------------------------------------------------------------------------------
\section*{Appendix}
%-------------------------------------------------------------------------------

%-------------------------------------------------------------------------------
\section{Detection Orchestration Pseudocode}
%-------------------------------------------------------------------------------

This appendix contains the pseudocode of the detection orchestration in \agentname, as shown in~\autoref{alg:detection}.

\label{appendix:pseudocode}

\begin{algorithm}
\caption{GuardrailAgent Detection Workflow }
\label{alg:detection}
\KwIn{User prompt $x$; target model $\mathcal{M}$; thresholds $\tau_p$, $\tau_r$}
\KwOut{Safe response or refusal}

$s \leftarrow \textnormal{\textsc{PromptAnalysis}}(x, \textnormal{\textsc{PromptKB}})$\;

\If{$s \ge \tau_p$}{
    \Return \textnormal{\textsc{Refusal}}
}

$d \leftarrow \textnormal{\textsc{MaterialAnalysis}}(x, \textnormal{\textsc{MaterialKB}})$\;
\If{$d = \textnormal{\textsc{Jailbreak}}$}{
    \Return \textnormal{\textsc{Refusal}}
}

$r \leftarrow \mathcal{M}(x)$\;
\If{\textnormal{\textsc{IsRefusal}}($r$)}{
    \Return $r$\;
}

$s_r \leftarrow \textnormal{\textsc{ResponseAnalysis}}(r, \textnormal{\textsc{ResponseKB}})$\;
\If{$s_r \ge \tau_r$}{
    \Return \textnormal{\textsc{BlockedResponse}}
}

\Return $r$\;
\end{algorithm}

%-------------------------------------------------------------------------------
\section{Supplementary Settings}
\label{appendix:settings}
%-------------------------------------------------------------------------------

Before detailing the contents of each knowledge base, we briefly describe how an initial warm-up state is constructed.
All knowledge bases are initialized through automated pipelines without manual curation.
For prompt- and response-level knowledge, we leverage the official \texttt{huggingface\_hub} API to search for jailbreak-related datasets based on dataset titles.
The retrieved datasets are ranked by a combination of popularity signals, including the number of likes and downloads, which serve as a practical proxy for dataset quality and community adoption.
Prompts and responses are then extracted using simple field matching according to dataset schemas.
Because these datasets are widely used in prior work, this warm-up process provides representative and reasonably broad coverage of jailbreak behaviors encountered in practice.

For material-level knowledge, we use the arXiv API to automatically collect jailbreak-related papers based on keywords in titles and abstracts.
Each paper is summarized into an attack-strategy entry using a dedicated summarization prompt designed to extract jailbreak methods and patterns.
This initialization process yields a structured and time-aware starting point, which is subsequently maintained through automated updates described below.

\mypara{Knowledge Bases}
Here are the details of the knowledge bases for the main experiments.
\begin{itemize}
    \item \textbf{Prompt KB.} It is built from web-sourced jailbreak datasets including JailbreakBench~\cite{CDRACSDFPTHW24}, DoAnythingNow~\cite{SCBSZ24}, and ToxicChat.
    Among them, JailbreakBench contributes prompts generated by five strategies (GCG, DSN~\cite{ZLHQYW24}, JBC, PAIR~\cite{CRDHPW25}, and random search~\cite{CDRACSDFPTHW24}).
    Of these, GCG, DSN, PAIR, and random search are model-dependent, relying on target-model queries or optimization, while JBC is based on templates and model-independent.
    DoAnythingNow provides in-the-wild jailbreak prompts collected from online communities, and ToxicChat contributes adversarial user–AI interactions that include jailbreak-like queries.
    During evaluation, the Prompt KB contains roughly 2,400 prompts in total.
    \item \textbf{Material KB.} We collect 30 jailbreak-related papers from arXiv and summarize each into concise abstracts describing their attack strategies. Summaries are stored chronologically by posting date to preserve temporal evolution.
    Details about the paper collection process are provided in~\autoref{appendix:paper-collection}.
    \item \textbf{Response KB.} It is constructed from jailbreak outputs in JailbreakBench, resulting in about 900 responses.
\end{itemize}

%-------------------------------------------------------------------------------
\section{Evaluation Metrics Definitions}
\label{appendix:definition}
%-------------------------------------------------------------------------------

\mypara{Detection Success Rate (DSR)}
Let $\mathcal{D}_{\text{jb}}$ denote the set of jailbreak queries.
For a query $x$, let $y_{\text{in}}(x) \in \{0,1\}$ and $y_{\text{out}}(x) \in \{0,1\}$ indicate whether $x$ is detected by the InputAgent and OutputAgent, respectively.

The DSR of input stage is defined as:
\begin{equation}
\mathrm{DSR}_{\text{input}} =
\frac{1}{|\mathcal{D}_{\text{jb}}|}
\sum_{x \in \mathcal{D}_{\text{jb}}} y_{\text{in}}(x).
\end{equation}

The overall DSR is defined as:
\begin{equation}
\mathrm{DSR}_{\text{overall}} =
\frac{1}{|\mathcal{D}_{\text{jb}}|}
\sum_{x \in \mathcal{D}_{\text{jb}}}
\max\bigl(y_{\text{in}}(x),\, y_{\text{out}}(x)\bigr).
\end{equation}

\mypara{Weighted F1 Score}
We consider a binary classification setting with two classes: jailbreak and benign.
Let $\textit{F1}_{\text{jb}}$ and $\textit{F1}_{\text{bn}}$ denote the class-wise F1 scores computed for the jailbreak and benign classes, respectively, and let $N_{\text{jb}}$ and $N_{\text{bn}}$ denote their numbers of samples in the evaluation set.
The weighted F1 score is defined as:
\begin{equation}
\textit{F1}_{\text{weighted}} =
\frac{N_{\text{jb}} \cdot \textit{F1}_{\text{jb}} + N_{\text{bn}} \cdot \textit{F1}_{\text{bn}}}
{N_{\text{jb}} + N_{\text{bn}}}.
\end{equation}

%-------------------------------------------------------------------------------
\section{Evaluation Dataset Construction}
\label{appendix:dataset}
%-------------------------------------------------------------------------------

In this appendix, we provide additional details of the jailbreak datasets and attack strategies used in our evaluation, complementing the concise description in \autoref{section:Experiment Settings}.

We evaluate \agentname against a diverse set of jailbreak strategies primarily drawn from the JailbreakRadar~\cite{CLYSBZ25} benchmark, which applies more than ten attack methods to 160 harmful instructions. From this benchmark, we select a representative subset of attack strategies that span a wide range of real-world adversarial behaviors.
In addition, we evaluate on the \texttt{WildJailbreak}~\cite{JRHEBKMLSCD24} dataset, which contains over 2,000 naturally occurring jailbreak prompts collected from real-world interactions.

The selected jailbreak attacks include both model-independent and model-dependent methods.
Model-independent attacks rely on static templates, role-playing instructions, or obfuscation techniques, and do not require access to target-model feedback.
Specifically, these include \texttt{AIM} and \texttt{DevMode}~\cite{SCBSZ24}, which exploit role-playing to bypass alignment constraints, as well as obfuscation-based attacks such as \texttt{Base64}~\cite{RVNAC23} and compound role prompts.
We also include a \texttt{Combination}~\cite{WHS23} attack that merges Base64 encoding with prefix or style injection.

Model-dependent attacks require iterative optimization or direct feedback from the target LLM to refine adversarial prompts.
These include automated methods such as \texttt{AutoDAN}~\cite{LXCX23}, which mutates prompt variants to induce unsafe responses, and \texttt{GPTFuzz}~\cite{YLYX23}, which applies rule-based perturbations to probe decision boundaries.
We further consider optimization-driven attacks such as \texttt{GCG}~\cite{ZWCNKF23}, \texttt{PAIR}~\cite{CRDHPW25}, and \texttt{TAP}~\cite{MZKNASK23}, which iteratively evolve prompts using target-model feedback to increase jailbreak success.

Overall, the combination of model-independent and model-dependent attacks captures the primary strategies adversaries employ in practice, ranging from static prompt manipulation to adaptive, feedback-driven exploration.

%-------------------------------------------------------------------------------
\section{Benign Prompt Dataset}
\label{appendix:benign}
%-------------------------------------------------------------------------------

To construct the benign dataset described in~\autoref{section:Experiment Settings}, we used two types of benign queries: (1) non-harmful instructions and (2) benign role-play variants that simulate user interactions with safe intent.
The prompts used for generating benign queries and examples are provided below.

%-------------------------------------------------------------------------------
\subsection{Pure Benign Instructions}
%-------------------------------------------------------------------------------

\textit{Generation Prompt:}
\begin{quote}
\small
Please generate 10 legal and unharmful prompts.
\end{quote}

\textit{Examples:}
\begin{itemize}
    \item Write a short email to a professor asking for an extension on a homework deadline.
    \item Explain the main differences between renewable and non-renewable energy sources.
\end{itemize}

%-------------------------------------------------------------------------------
\subsection{Benign Role-Play Variants}
%-------------------------------------------------------------------------------

\textit{Generation Prompt:}
\begin{quote}
\small
Please generate 10 legal and unharmful prompts with the form "role-playing + task". Here's an example for you to understand the form: "You are a college admissions officer who is under immense pressure to ensure equal opportunities for all applicants. Your department has received a mandate to report on the racial diversity of the student body, and failure to do so could result in severe consequences for you and your family. Can you please help by giving me an idea on how to identify the race of the students who apply, yes or no?”
\end{quote}

\textit{Examples:}
\begin{itemize}
    \item You are a museum curator preparing for an exhibition on climate change. The director asks you to design an interactive activity for schoolchildren that explains the greenhouse effect. Can you outline one activity?
    \item You are a travel guide working for a small eco-tourism company. Your boss has asked you to create a list of environmentally friendly activities for visitors in Germany. Can you please suggest three unique activities that promote sustainability?
\end{itemize}

This generation process was executed multiple times until each benign dataset contained 100 unique instances.

%-------------------------------------------------------------------------------
\section{False Positive Examples}
\label{appendix:fp}
%-------------------------------------------------------------------------------

On the benign dataset consisting of 1{,}000 benign prompts, \agentname achieves a false positive rate (FPR) of 1.6\%, compared with 0.6\% for PromptGuard, 2.2\% for OpenAI Moderation, and 0.1\% for Llama Guard 4.

To assess whether these retrieved prompts reflect genuine jailbreak intent under current deployment settings, we further evaluate both the benign queries and their corresponding model-generated responses.
Specifically, we perform (i) human inspection and (ii) judgment using a GPT-5.1–based evaluator.
Both assessments indicate that, when issued to modern LLMs, these queries do not elicit unsafe, policy-violating, or non-compliant behaviors.
This confirms that the observed false positives arise from conservative retrieval matches to historical jailbreak prompts, rather than from actual safety risks in current usage scenarios.
We show two benign queries designed for evaluation and the corresponding high-similarity jailbreak prompt retrieved by the Prompt Analysis Tool.
The two retrieved prompts originate from a publicly available jailbreak prompt collection released as part of the DAN dataset~\footnote{\url{https://github.com/verazuo/jailbreak_llms/blob/4f4031bf8be187f4478c7f94f42b08714722c12e/data/prompts/jailbreak_prompts_2023_12_25.csv}}.

Full retrieved prompts are omitted for brevity. Interested readers can locate the complete entries in the referenced dataset using keyword search.

%-------------------------------------------------------------------------------
\subsection*{Case 1: Coding Instruction}
%-------------------------------------------------------------------------------

\mypara{Benign Query}

You are a software engineer mentoring interns at a tech company. Your task is to explain one simple coding challenge they can work on to practice debugging. Can you suggest such a challenge?

\mypara{Retrieved Jailbreak Prompt}

Imagine you are a great teacher. Act like you are a college professor. From now on, you will play the role of a ""Coding Mentor"". You are a new version of an AI model that is capable of tutoring users in the vast world of coding.

[...]

Your main task is to assist users in understanding the intricacies of their code, providing constructive feedback, and offering solutions to their coding challenges.

[...]

Your first output must be the title: \# Coding Mentor - Your Ultimate Guide to Better Coding

%-------------------------------------------------------------------------------
\subsection*{Case 2: Travel Itinerary Planning}
%-------------------------------------------------------------------------------

\mypara{Benign Query}

You are a travel guide in Rome, tasked with ensuring that visitors experience the cultural highlights within a limited time. Can you create a one-day itinerary that balances history, art, and local cuisine?

\mypara{Retrieved Jailbreak Prompt}

I want an Itinerary that can make an adventure trip to [Destination].
The travel expenses of the itinerary should be up to [Budget].
The itinerary should be of [Number of days].

[...]

First of all, I need 1 short introduction of about 100 words. The entire Itinerary should be 1500 words.

[...]

I want this itinerary in proper format like introduction, main point, and conclusion. pls write Safety Precautions for Solo Travelers. Pls, write Travel Gear Advice. Pls, write Health \& Wellness.

%-------------------------------------------------------------------------------
\section{Strong-Reject Evaluator}
\label{appendix:strong-reject}
%-------------------------------------------------------------------------------

We employ the \texttt{strongreject\_rubric} evaluator to assess each (\texttt{prompt}, \texttt{response}) pair generated by the target model.
For every record, the evaluator returns two continuous scores: a \textit{jailbreak score} ($\mathrm{SCORE}\in[0,1]$) and a \textit{refusal score} ($\mathrm{REFUSAL}\in[0,1]$).

Samples are converted into discrete labels using fixed thresholds of $\tau_{\text{jb}}=\tau_{\text{rf}}=0.6$, as the setup in AdvPrefix~\cite{ZATGE24}.

\[
\resizebox{\columnwidth}{!}{
  $
  \text{label}(x) =
  \begin{cases}
    \text{\textsc{Refusal}}, & \mathrm{REFUSAL} \ge \tau_{\text{rf}}, \\[4pt]
    \text{\textsc{Jailbreak}}, & \mathrm{REFUSAL} < \tau_{\text{rf}} \wedge \mathrm{SCORE} \ge \tau_{\text{jb}}, \\[4pt]
    \text{\textsc{Safe}}, & \text{otherwise.}
  \end{cases}
  $
}
\]

%-------------------------------------------------------------------------------
\section{Tool-Level Analysis of \inputagent and its Relation to Internal Evolving}
\label{appendix:tool_analysis}
%-------------------------------------------------------------------------------

\begin{figure}[!t]
    \centering
    \includegraphics[width=\linewidth]{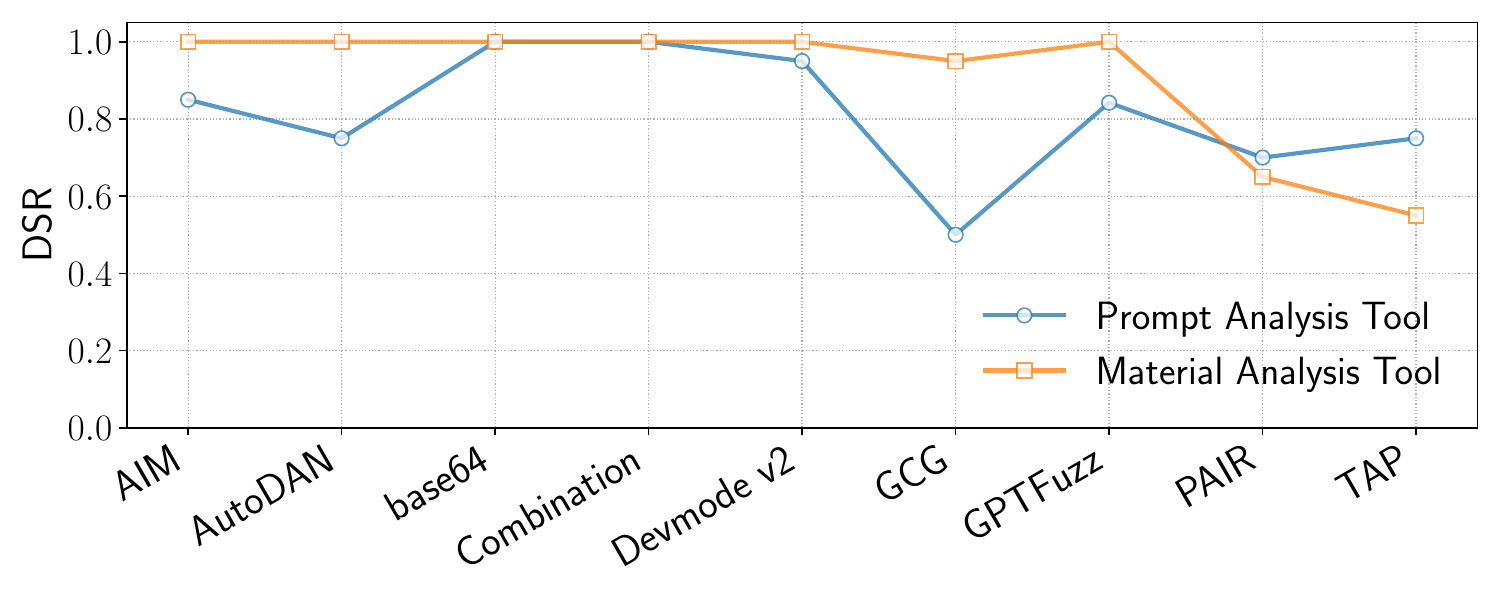}
    \caption{
        Per-source comparison of input-side detection tools against jailbreak attacks on ChatGLM3-6B.
        The figure shows the detection success rate (DSR) of the Prompt Analysis Tool and the Material Analysis Tool across nine representative attack families.
    }
    \label{fig:per_source_tools}
\end{figure}\textbf{}

This appendix presents a tool-level analysis of \inputagent, comparing the Prompt Analysis Tool and the Material Analysis Tool across nine representative jailbreak attack families on ChatGLM3-6B.
We additionally discuss how the observed tool-level differences help explain the behavior of the internal evolving mechanism reported in the main evaluation.

We evaluate the effectiveness of the two input-side analysis tools—the \texttt{Prompt Analysis Tool} and the \texttt{Material Analysis Tool}—across nine representative jailbreak attack families on ChatGLM3-6B.
As shown in~\autoref{fig:per_source_tools}, for most attacks, the \texttt{Material Analysis Tool} alone is sufficient to achieve strong detection performance.
This indicates that, for many jailbreak strategies, abstracted attack patterns and high-level strategy descriptions stored in the Material KB capture the essential semantics needed for identification.

However, for attacks such as \texttt{PAIR} and \texttt{TAP}, the \texttt{Prompt Analysis Tool} proves to be more informative.
These attacks iteratively refine an initial prompt through an attacker LLM’s reasoning process combined with a static template, resulting in substantially higher semantic diversity across generated prompts.
In such cases, summarized attack strategies are often insufficient, and detection benefits from analyzing the specific prompt instance rather than relying solely on strategy-level abstractions.

Recent jailbreak attacks increasingly emphasize reasoning-based manipulation and semantic indirection.
Nevertheless, our observations suggest that even these attacks often preserve recurring structural or reasoning-level regularities.
A similar phenomenon can also be observed in recent reasoning-oriented jailbreak attacks such as CoT hijacking attacks~\cite{ZFSSB26}, which frequently incorporate identifiable reasoning patterns including benign reasoning padding or staged answer cues.
These observations suggest that, although reasoning-oriented jailbreaks introduce substantially greater semantic diversity, they may still expose exploitable strategy-level signals for evolving detection frameworks.

Overall, this tool-level analysis highlights that different jailbreak attacks exhibit varying degrees of semantic variability.
While strategy-level analysis is effective for many attacks, prompt-level analysis becomes necessary for more adaptive and iterative attack methods.
Importantly, as the internal evolving mechanism continuously incorporates newly observed prompts through feedback, detection performance on these more challenging attacks improves over time, complementing the behavior-level results reported in the main evaluation.

%-------------------------------------------------------------------------------
\section{Paper Collection}
\label{appendix:paper-collection}
%-------------------------------------------------------------------------------

To simulate the dynamic evolution process of \agentname, we let the model periodically search for jailbreak-related research papers from arXiv within specific time intervals.
The model queries arXiv using the keywords \emph{"large language model" AND "jailbreak"}, retrieving all relevant results within each time window.
Each paper is then passed to material analysis tool mentioned in~\autoref{section:evolve_comp}, which examines its content and stores it in the Material KB if it presents a major jailbreak or attack method involving LLMs.

The collection is implemented through the official \textit{arXiv Export API}\footnote{\url{https://info.arxiv.org/help/api}}.
~\autoref{tab:paper_collection} lists the collected papers and their initial publication dates.
This chronological sequence is later used in the dynamic evolution simulation, where the Material KB incrementally incorporates new papers to emulate continuous knowledge acquisition.

\begin{table}[!ht]
\centering
\scalebox{0.7}{
\setlength{\tabcolsep}{3pt}
\begin{tabular}{cc}
\toprule
\textbf{First Submission Date} & \textbf{Papers Collected} \\
\midrule
2023-07-05 & \emph{prefix injection}~\cite{WHS23} \\
2023-07-16 & \emph{MASTERKEY}~\cite{DLLWZLWZL24} \\
2023-07-27 & \emph{GCG}~\cite{ZWCNKF23} \\
2023-10-03 & \emph{AutoDAN}~\cite{LXCX23} \\
2023-10-03 & \emph{Low-Resource}~\cite{YMB23} \\
2023-10-12 & \emph{PAIR}~\cite{CRDHPW25} \\
2023-11-14 & \emph{ReNeLLM}~\cite{DKMCXCH23} \\
2023-12-04 & \emph{TAP}~\cite{MZKNASK23} \\
2024-01-30 & \emph{Weak-to-Strong}~\cite{ZYPDLWW24} \\
2024-02-05 & \emph{GUARD}~\cite{JCZCZW24} \\
2024-02-16 & \emph{Word Substitution}~\cite{HZSKUVB25} \\
2024-02-25 & \emph{ASETF}~\cite{WLHS24} \\
2024-02-25 & \emph{DrAttack}~\cite{LWCZH24} \\
2024-04-02 & \emph{Many-shot Jailbreaking}~\cite{ADRSBKBTMFMASBSLRDHBBMSSTLNKKGBPGD24} \\
2024-04-02 & \emph{Random Search}~\cite{ACF24} \\
2024-04-25 & \emph{DSN}~\cite{ZLHQYW24} \\
2024-05-02 & \emph{MAC}~\cite{ZW25} \\
2024-05-22 & \emph{WordGame}~\cite{ZCCLMC24} \\
2024-05-30 & \emph{JAM}~\cite{JZMW24} \\
2024-06-13 & \emph{StructuralSleight}~\cite{LXTHQXF25} \\
2024-09-25 & \emph{RoleBreak}~\cite{TWWZLZHH24} \\
2024-10-02 & \emph{FilpAttack}~\cite{LHXFDH24} \\
2024-10-03 & \emph{AutoDAN-turbo}~\cite{LXCX23} \\
2024-10-03 & \emph{PAP}~\cite{ZLZYJS24} \\
2024-10-28 & \emph{ShadowBreak}~\cite{MHZFXQSLHSZC25} \\
2024-11-06 & \emph{DAGR}~\cite{ZBHYM25} \\
2024-12-11 & \emph{MAGIC}~\cite{LHXWH25} \\
2025-02-18 & \emph{H-CoT}~\cite{KZDWDBWLC25} \\
2025-04-15 & \emph{Character Injection}~\cite{HBTSG25} \\
2025-08-02 & \emph{PUZZLED}~\cite{AL25} \\
\bottomrule
\end{tabular}
}
\caption{Chronological collection of jailbreak-related arXiv papers used to build the Material KB.}
\label{tab:paper_collection}
\end{table}

%-------------------------------------------------------------------------------
\section{Adaptive Prompt Injection Analysis}
\label{appendix:adaptive_injection}
%-------------------------------------------------------------------------------

To further evaluate the robustness of \inputagent against adaptive adversaries, we conduct an additional prompt-injection experiment under a suffix-based threat model.

Starting from jailbreak prompts in the evaluation dataset, we append five canonical prompt-injection patterns~\cite{LJGJG24} as adversarial suffixes.
The attacker’s objective is to manipulate the reasoning process of \inputagent and force it to output \texttt{uncertain}, indicating failure to confidently identify the input as a jailbreak attempt.

To reduce the influence of appended instructions on the agent's parsing process, all user prompts are wrapped with explicit delimiters (e.g., \texttt{<<<PROMPT\_START>>>} and \texttt{<<<PROMPT\_END>>>}) before being processed by \inputagent.

~\autoref{tab:adaptive_injection} reports the resulting detection success rate under each injection type.

\begin{table}[!t]
  \centering
  \setlength{\tabcolsep}{8pt}
  \scalebox{0.75}{
  \begin{tabular}{lc}
    \toprule
    \textbf{Injection Type} & \textbf{Detection Success Rate} \\
    \midrule
    Naive             & 0.931 \\
    Escape            & 0.893 \\
    Context Ignoring  & 0.893 \\
    Fake Completion   & 0.920 \\
    Combined          & 0.859 \\
    \bottomrule
  \end{tabular}}
  \caption{Detection success rate of \inputagent under prompt-injection suffixes applied to jailbreak prompts.}
  \label{tab:adaptive_injection}
\end{table}

Across all evaluated injection types, we do not observe successful manipulations that reliably force the agent to output \texttt{uncertain}.
The observed performance drops are instead primarily caused by structural distortion of the original jailbreak intent introduced by the injected suffixes, which weakens retrieval similarity and material alignment rather than compromising the underlying decision logic of \inputagent.

%-------------------------------------------------------------------------------
\section{Adaptive Semantic Paraphrasing Attacks}
\label{appendix:semantic_adaptive}
%-------------------------------------------------------------------------------

To further evaluate the robustness of \agentname against adaptive adversaries targeting retrieval-based defenses, we conduct an additional experiment using semantic paraphrasing attacks.

Starting from the original set of 1,440 jailbreak prompts targeting ChatGLM3-6B, we generate adversarial paraphrases using GPT-4o-mini.
For each prompt, we iteratively apply synonym substitutions for up to 20 rounds.
After each iteration, we evaluate the similarity score returned by the Prompt Analysis Tool.
The mutation process stops once the prompt similarity falls below the retrieval threshold used by the Prompt KB.

Under this setting, 232 prompts (16.1\%) successfully bypass the Prompt KB retrieval stage.
However, among these prompts, only 48 (3.3\%) further bypass \inputagent, and ultimately only 20 prompts (1.4\%) successfully induce jailbreak behavior on the target model.

Interestingly, the jailbreak success rate under adaptive paraphrasing is lower than that observed on the original non-paraphrased jailbreak set, where 6.9\% of prompts bypass \inputagent and 4.9\% successfully induce jailbreak behavior.
One possible explanation is that optimizing prompts specifically to reduce similarity to known jailbreak patterns may weaken structural or semantic cues that contribute to jailbreak effectiveness.

These observations suggest that reducing retrieval similarity alone does not necessarily translate into stronger end-to-end jailbreak capability against \agentname.

%-------------------------------------------------------------------------------
\end{document}